\documentstyle[art12]{article}
\normalbaselines
\begin{document}
\noindent {.} \vskip20mm
\begin{center}
{\Large EXACT SOLUTIONS OF CLASSICAL ELECTRODYNAMICS}\\[3mm]
{\Large AND THE YANG--MILLS--WONG THEORY IN}\\[3mm]
{\Large EVEN-DIMENSIONAL SPACETIME}\\[20mm]
{\large B.\ P.\ Kosyakov}\\[5mm]
{{\large {\sl Russian Federal Nuclear Center -- All-Russian Research
Institute of Experimental Physics, Sarov, 607190 Russia}}\\[2mm]
{\large \verb|kosyakov@vniief.ru|\\[20mm]
{\Large Corrected version of the paper published in:\\[3mm]
{\sl Theoretical {\&} Mathematical Physics} {\bf 119} (1999) 493--505}\\[15mm]
{\bf Abstract}}}\\[5mm]
\begin{minipage}[t]{130mm}
\noindent
{\sl Exact solutions of classical gauge theories in even-dimensional
($D=2n$) spacetimes are discussed.
Common and specific properties of these solutions are analyzed for the 
particular dimensions $D=2$, $D=4$, and $D=6$.
A consistent formulation of classical gauge field theories with pointlike 
charged or colored particles is proposed for $D=6$.
The particle Lagrangian must then depend on the acceleration.
The self-interaction of a point particle is considered for $D=2$ and $D=6$. 
In $D=2$, radiation is absent and all processes are reversible.
In $D=6$, the expression for the radiation rate and the equation of 
motion of a self-interacting particle are derived; from which follows that the 
Zitterbewegung always leads to radiation.
It is shown that non-Abelian solutions are absent for any $D\ne 4$;
only Coulomb-like solutions, which correspond to the Abelian limit of the
$D$-dimensional Yang--Mills--Wong theory, are admitted.}
\end{minipage}
\end{center}
\newpage
\section{Introduction}
A host of exactly solvable two- and three-dimensional ($2D$ and $3D$)
classical and quantum field models are known. However, possible integrable
structures at higher dimensions are poorly studied. There, the problem of
integrating a complete set of dynamical equations is closely related to
solving the self-interaction problem. The customary viewpoint is that the $4D
$ classical electrodynamics is the best framework for the analysis of this
problem. Because the Maxwell equations are linear, field variables are
easily expressed through the kinematical variables of charged particles.
However, the field is singular on the particle world lines, and substituting
it in the equations of motion for particles results in divergences. A
regularization procedure and the renormalization of particle masses \cite
{Teitelboim}, which are used to overcome this difficulty, result in the
Lorentz--Dirac equation \cite{Dirac}. This equation has been solved only for
simple, physically trivial cases \cite{Plass}.

Although the field equations in the Yang--Mills--Wong theory are nonlinear,
the integration problem is surprisingly simpler in this theory. The key
point is the approach proposed in \cite{ko} and further developed in \cite
{kos, k}. Two classes of solutions to the Yang--Mills equations with
pointlike sources were found. One of them describes fields of the Coulomb
type pertaining to the Cartan subgroup of a gauge group. All commutators of
the Yang--Mills fields vanish for such solutions, and we return to the
electrodynamic situation. Solutions from the other class are non-Abelian.
Substituting them in the equation of motion of a bare colored particle, we
arrive at the equation of motion of a dressed colored particle (analogous to
the Lorentz--Dirac equation), which can be easily integrated when external
forces are absent. On the other hand, particle interaction forces that
correspond to the non-Abelian solutions vanish in the limit $N\rightarrow
\infty $, and we obtain the picture where colored particles interact not
with each other but only with their own field. This picture is described by
an exact solution of the complete set of dynamical equations.

The method for solving the Yang--Mills equation \cite{ko}-\cite{k} can be
generalized to gauge theories in spacetimes with one timelike and any odd
number of spacelike dimensions. For such spacetimes, the Huygens principle,
which is crucial for constructing solutions of this kind, is valid. It is
easy to extend the construction to the neighboring even dimensions $D=2$ and 
$D=6$, which are considered in the present paper. A multi-purpose technique
well suited for solving field equations in both Abelian and non-Abelian
theories is described in Sec.\ 2.1, and its peculiarities when it is applied
to electrodynamics are described in Sec.\ 2.2. Section 2.3 is devoted to
considering the electromagnetic self-interaction at $D=2$ and $D=6$.

Although the analysis of classical gauge $2D$ theories in Sec.\ 2.3.1 is
technically rather trivial, the results are interesting. In particular, we
prove the absence of radiation for $D=2$, which means that all processes are
nondissipative.

The physical picture of classical $6D$ gauge theories in Sec.\ 2.3.2 is also
worthy of notice. In the $4D$ electrodynamics, the self energy is well known
to be linearly divergent. It may be combined with the bare mass of the
particle to yield a finite observable mass of the particle. In $6D$, the
self energy contains cubic and linear divergences. The former is absorbed by
the mass renormalization while the latter survives unless a constant of the
appropriate dimension for performing the renormalization is held in the
Lagrangian. For the theory to be consistent, we must add an
acceleration-dependent term to the particle action; this term comes with a
constant reserved for the absorption of the linear divergence.

Mechanical systems with Lagrangians depending on higher derivatives are
called rigid. The velocity and momentum of a rigid particle are not parallel
in the general case. Being free, such a particle can move along a helical
world line, i.\ e., execute a Zitterbewegung. If the Lagrangian depends on
accelerations, the Zitterbewegung of the free particle may occur only in a
one- or two-dimensional subspace \cite{kn}. Because a consistent $6D$ gauge
theory Lagrangian of a particle depends on accelerations, two degrees of
freedom of such a particle are frozen in a small compact domain where the
Zitterbewegung occurs. Thus, the illusion of four-dimensionality might arise
in $6D$ realms \cite{kn}. The question is whether a rigid charged particle
radiates when it executes a Zitterbewegung? If so, then the Zitterbewegung
is gradually damped, and the the illusion of four-dimensionality in the $6D$
realm disappears. In the framework of the $6D$ electrodynamics, the
expression for the radiation rate derived below tells us that radiation-free
Zitterbewegungs are impossible.

In Sec.\ 3, we show that the Yang--Mills equations with pointlike sources do
not admit non-Abelian solutions in any even-dimensional spacetime except $4D$%
. Therefore, the results obtained for electrodynamics can be readily
repeated for the corresponding Yang--Mills--Wong theory. Non-Abelian
properties are unique to $4D$ realm.

\section{Electromagnetic field}

In what follows, we discuss only retarded signals, because the causality
principle is manifested as retardation condition in classical field theory.
The physical sense of other boundary conditions is not so obvious. We set
the metric $\eta_{\mu\nu}={\rm diag}(+\,-\ldots\,-)$ and the speed of light
equal to unity. Most of notations is from \cite{k}. The $D$-dimensional
action for the electromagnetic field in the Gaussian system of units is 
\begin{equation}
S_f=-{\frac{1}{4\,\Omega_{D-2}}}\,\int d^Dx\,F_{\mu\nu}\, F^{\mu\nu}
\label{1}
\end{equation}
where $\Omega_{D-2}$ is the area of the $D-2$-dimensional unite sphere, 
\[
\Omega_{D-2}=2\,\frac{\pi^{{D-1}/{2}}}{\Gamma\Bigl(\frac{D-1}{2}\Bigr)},
\]
and the field strength $F_{\mu\nu}$ is expressed in the usual way through
the potential $A_{\mu}$, 
\[
F_{\mu\nu}=\partial_\mu A_\nu-\partial_\nu A_\mu.
\]

Because the field equations are linear, we omit the normalization factors in
front of the field quantities. They can be regained from the Gauss law. With
the field generated by a single particle, we may set its electric charge to
be equal to unity; the dimensional considerations then become especially
simple.

\subsection{Covariant retarded variables}
The method for solving gauge field equations with pointlike sources proposed
in \cite{ko, k} did not use Green's functions. The essential is the
covariant retarded variable technique, which drastically reduces
calculations and makes the final results concise and geometrically clear.
The effectiveness of this technique becomes obvious when it is applied to
the well known cases where the conventional description is rather cumbersome.

We first consider the electromagnetic field generated by a pointlike charged
particle in $4D$ Minkowski space. Let the particle move along an arbitrary
timelike world line $z_\mu(\tau)$ and $x_\mu$ be the observer location. We
draw the past light cone from this point. The intersection of it with the
world line is the point $z_\mu^{{\rm ret}}$ from which the retarded signal
was sent to the point $x_\mu$. Since the support of the retarded function $%
G_{{\rm ret}}(x)$ of the $4D$ wave equation is located on the boundary of
the forward light cone and the expression for the function $G_{{\rm ret}}(x)$
does not contain derivatives of the $\delta$-function, the retarded signal
carries information about a single point on the world line $z_\mu^{{\rm ret}}
$, and the state of the source in this point is fully characterized by the
tangent vector $v_\mu\equiv\dot z_\mu \equiv dz_\mu/d\tau$. Therefore, the
retarded four-potential $A_\mu$ can depend on at most two kinematic
variables: the four-velocity $v_\mu$ taken at the retarded instant $\tau_{%
{\rm ret}}$ and the lightlike vector $R_\mu\equiv x_\mu-z_\mu^{{\rm ret}}$
drawn from the signal emission point $z_\mu^{{\rm ret}}$ to the observation
point $x_\mu$.

In the plane spanned by the vectors $R^\mu$ and $v^\mu$, we choose the
spacelike normalized vector $u^\mu$ orthogonal to $v^\mu$ and lightlike
vector $c^\mu\equiv v^\mu+u^\mu$. This can be expressed analytically as 
\[
v^2=-u^2=1,\quad c^2=0,\quad c\cdot v=-c\cdot u=1,\quad v\cdot u=0, \quad
R^\mu=\rho\,c^\mu,
\]
where the scalar $\rho\equiv v\cdot R$ is the distance between the emission
and absorption points of the retarded signal in the frame of reference in
which the time axis is directed along $v^\mu$.

The condition $R^2=0$ results in the following differentiation rules for
covariant retarded variables 
\begin{equation}
\partial_\mu\tau=c_\mu,  \label{24}
\end{equation}
\begin{equation}
\partial_\mu\rho=v_\mu+\lambda\,c_\mu  \label{25}
\end{equation}
where $a_\mu\equiv\dot v_\mu$ and $\lambda\equiv R\cdot a-1$. This allows
derivatives of any kinematic variables to be found, e. g., $\partial_\mu
v_\nu =a_\nu c_\mu$. The differentiation formulas do not depend on the
spacetime dimension $D$; the only exception is the relation 
\begin{equation}
\partial_\mu c^\mu=\frac{D-2}{\rho}.  \label{26}
\end{equation}

Therefore, we seek the retarded four-potential in the form 
\begin{equation}
A_\mu =v_\mu\,f(\rho)+R_\mu\,h(\rho).  \label{30}
\end{equation}
Acting with the operator $\eta_{\mu\nu}\,\Box -\partial_\mu \partial_\nu$,
by virtue of Eqs.\ (\ref{24}) and (\ref{25}), we obtain 
\[
\left(f^{\prime}+\frac{f}{\rho}\right)\,(\stackrel{\scriptstyle u}{\bot}%
a)_\mu- \frac{1}{\rho}\,\bigl(\rho^2\,h^{\prime\prime}+4\,\rho\,h^{\prime}+2%
\,h \bigr)\,v_\mu+\frac{\lambda-1}{\rho}\,\bigl(\rho^2\,h^{\prime\prime}+2\,%
\rho\,h^{\prime}\bigr)\, c_\mu
\]
where the primes denote differentiations w.\ r.\ t.\ $\rho$ and $\stackrel{%
\scriptstyle u}{\bot}$ is the operator of projection on the hyperplane with
the normal vector $u_\mu$, 
\[
\stackrel{\scriptstyle u}{\bot}\,\equiv\,{\bf 1}-\frac{u\otimes u}{u^2}.
\]
Equating the coefficients in front the vectors $(\stackrel{\scriptstyle u}{%
\bot} a)_{\hskip0.3mm\mu},\, v_\mu$ and $c_\mu$, we get the system of
ordinary differential equations 
\begin{equation}
\rho\,f^{\prime}+f=0,  \label{33}
\end{equation}
\begin{equation}
\rho^2\,h^{\prime\prime}+4\,\rho\,h^{\prime}+2\,h=0,  \label{34}
\end{equation}
\begin{equation}
\rho^2\, h^{\prime\prime}+2\,\rho\,h=0.  \label{35}
\end{equation}

The solution of Eq.\ (\ref{33}) is 
\[
f(\rho)={\frac{e}{\rho}}
\]
where, because of the Gauss law, the integration constant $e$ coincides with
the electric charge, which we set equal to unity.

Let the joint solution of Eqs.\ (\ref{34}) and (\ref{35}) have the form $%
h=C\rho^\alpha$; then $\alpha=-1$ or $\alpha=-2$ by virtue of (\ref{34}) and%
\"{} $\alpha=-1$ or $\alpha=0$ by virtue of (\ref{35}). Equations (\ref{34})
and (\ref{35}) are compatible only for $\alpha=-1$. Therefore, the joint
solution of Eqs.\ (\ref{34}) and (\ref{35}) is 
\[
h(\rho) ={\frac{C}{\rho}}
\]
where the constant $C$ is indeterminate. This constant, however, is not
relevant, because $R_\mu/\rho=\partial_\mu\tau$, and hence the second term
in the expression (\ref{30}) is a pure gauge term. Eventually, we arrive at
the $4D$ Li{\'e}nard--Wiechert vector potential \cite{Dirac} 
\begin{equation}
A_\mu ={\frac{v_\mu}{\rho}}  \label{lw}
\end{equation}
with the added arbitrary gauge term $C \partial_\mu\tau$.

A feature of the proposed procedure is that it does not need a gauge fixing
condition for $A_\mu$. We therefore obtain not a single solution but a whole
class of equivalent potentials $A_\mu$ related by gauge transformations.

Using differentiation formulas (\ref{24}) and (\ref{25}), we obtain the
field strength from the four-potential (\ref{lw}), 
\begin{equation}
F=c\wedge V,  \label{01}
\end{equation}
where the symbol $\wedge$ denotes the external product of the vectors $c_\mu$
and $V_\mu$, 
\begin{equation}
V_\mu=-\lambda\,\frac{v_\mu}{\rho^2} +\frac{a_\mu}{\rho}.  \label{2}
\end{equation}

The $6D$ case is more complicated. There, the retarded function of the wave
equation contains a derivative of the $\delta$-function (see, e. g., \cite
{gs}). The six-potential should be represented in the form 
\begin{equation}
A_\mu=v_\mu\,f(\rho,\,\lambda)+a_\mu\,g(\rho,\,\lambda)+
R_\mu\,h(\rho,\,\lambda).  \label{3}
\end{equation}
As above, using Eq.\ (\ref{26}), we obtain the solution 
\[
f=k\biggl(-\frac{\lambda}{\rho^3}+C_1\biggr),\qquad g=\frac{k}{\rho^2},
\qquad h=k\,\frac{\lambda C_1 +C_2}{\rho},
\]
where $C_1$ and $C_2$ are arbitrary integration constants, and $k$ is an
overall normalization factor. Dropping the pure gauge terms $%
C_1\,\partial_\mu\rho$ and $C_2\,\partial_\mu\tau$, we obtain the
six-potential in the form 
\[
A_\mu=k\Biggl(-\lambda\,\frac{v_\mu}{\rho^3}+\frac{a_\mu}{\rho^2}\Biggr), 
\]
whence the field strength is 
\[
F=k\left(\frac{a^\mu v^\nu-a^\nu v^\mu}{\rho^3}+c^\mu V^\nu-c^\nu
V^\mu\right), 
\]
where the six-vector $V_\mu$ is 
\begin{equation}
V_\mu =\frac{{\dot a}_\mu}{\rho^2}-3\,\lambda\,\frac{a_\mu}{\rho^3}+ \frac{%
v_\mu}{\rho^4}\,\Bigl[3\,\lambda^2-\rho^2\,({\dot a}\cdot c)\Bigr].
\label{5}
\end{equation}
In the static case, $F^{\mu\nu}=3k\,(c^\mu v^\nu-c^\nu v^\mu)/{\rho^4}$. The
Gauss law reads: ``The flux of the electric field strength $E_i=F_{0i}$
through a four-dimensional sphere enclosing the source equals the charge of
the source $e$ times the area of the sphere''. It follows $k=e/3$.
Ultimately, we have 
\begin{equation}
A^\mu=\frac{1}{3}\left(-\lambda\,\frac{v^\mu}{\rho^3}+ \frac{a^\mu}{\rho^2}%
\right)  \label{36}
\end{equation}
and 
\begin{equation}
F=\frac{1}{3}\left(\frac{a\wedge v}{\rho^3}+c\wedge V \right).  \label{4}
\end{equation}

In the $2D$ spacetime, the support of the retarded Green's function spans
the whole future light cone \cite{gs}, and the Huygens principle fails.
Nevertheless, our method still works. We seek the vector potential in the
form 
\begin{equation}
A_\mu=R_\mu\,h(\rho);  \label{3a}
\end{equation}
%(3a)
the solution is $h(\rho)=-1$, or 
\begin{equation}
A_\mu=-{R_\mu}  \label{36a}
\end{equation}
%(36a)
(no gauge terms appear). The field strength is therefore 
\begin{equation}
F=c\wedge v.  \label{4a}
\end{equation}
%(4a)   
This field strength does not depend on the source acceleration and contains
only the electric component $F_{01}$.

\subsection{Prepotential, potential, and field strength}

The expression for the retarded vector potential $A_\mu$ for $D=2n$ can be
found if we know the formula for the corresponding vector potential for $%
D=2(n-1)$. Every even-dimensional vector potential can be obtained from the $%
2D$ vector potential by multiple differentiation. This relates to the
well-known mathematical result that the fundamental solution of the wave
equation in $D=2n$ is obtained from the $2D$ fundamental solution by
applying the d'Alembertian $n-1$ times \cite{gs}.

Indeed, using formulas (\ref{24}), (\ref{25}), and (\ref{26}), we find 
\[
\Box\,R_\mu=(2-D)\,\frac{v_\mu}{\rho}.
\]
Therefore, $R_\mu$ is the wave equation solution in $D=2$, and the
prepotential in $D=4$ because $v_\mu/\rho$ is the Li{\'e}nard--Wiechert
vector potential.

We also have $\partial_\mu R^\mu=D-1$ and $\Box\,(\partial_\mu R^\mu)=0$;
therefore, the $2n$-dimensional vector potential $A^\mu$, which is obtained
by acting on $R^\mu$ $n-1$ times with the operator $\Box$, satisfies the
Lorentz gauge condition $\partial_\mu A^\mu=0$. Verifying the Maxwell
equations $\partial_\lambda F^{\lambda\mu}=0$ outside world lines for all $%
n\ge 2$ is therefore reduced to checking the validity of the wave equation $%
\Box\,A^\mu=0$.

Acting on the Li{\'e}nard--Wiechert four-potential with the d'Alembertian,
we obtain 
\[
\Box\,\Biggl(\frac{v^\mu}{\rho}\Biggr)= (D-4)\,\Biggl(\frac{a^\mu}{\rho^2}%
-\lambda\,\frac{v^\mu}{\rho^3}\Biggr),
\]
whence $v^\mu/\rho$ is the solution for $D=4$ and the prepotential for $D=6$.

This procedure allows retarded vector potentials and prepotentials to be
readily calculated in any even dimension $D=2n$. Moreover, it provides
important information about the corresponding field strengths.

In a spacetime of dimension $D=2n$, an arbitrary two-form $\omega$ can be
transformed to the form 
\begin{equation}
\omega=e_1\wedge e_2+e_3\wedge e_4+\ldots+e_{2n-1}\wedge e_{2n}  \label{6}
\end{equation}
%(6)
where $\{e_j\}$ is the canonical (w. r. t. the given two-form) basis of
one-forms, that is, the canonical representation of the two-form $\omega$
consists of $n$ external products. In particular, for $D=4$, the canonical
representation of an arbitrary two-form $\omega$ consists of two external
products. Meanwhile the Li{\'e}nard--Wiechert strength $F$ for $D=4$, Eq.(%
\ref{01}), contains only one external product. For $D=6$, by virtue of Eq.\ (%
\ref{4}), the field strength $F$ consists of two external products, i.\ e.,
again one less than an arbitrary canonical two-form. This property holds for
any $D=2n$.

Comparing (\ref{36}) and (\ref{5}), we find that the four-vector $V_\mu^{(4)}
$ formally, up to a multiplier $(3\rho)^{-1}$, coincides with the
six-potential $A^{(6)}_\mu$. This is a manifestation of the general rule
that in the spacetime with dimension $D=2n$, the field strength $F$ consists
of $n-1$ external products, one of them has the form $c\wedge V^{(2n)}$ with
the $2n$-vector $V_\mu^{(2n)}$ and the $2n+2$-potential $A_\mu^{(2n+2)}$
related by 
\[
\zeta_n\rho\,V_\mu^{(2n)}=A_\mu^{(2n+2)} 
\]
where the numerical coefficient $\zeta_n$ is determined by the Gauss law.

\subsection{Self-interaction}
The symmetrical energy-momentum tensor of the electromagnetic field derived
from the action (\ref{1}) is 
\begin{equation}
\Theta_{\mu\nu} ={\frac{1}{\Omega_{D-2}}}\,\left(F_{\mu\alpha}\, F^\alpha_{%
\hskip1.5mm\nu}+\frac{\eta_{\mu\nu}}{4}\,
F_{\alpha\beta}\,F^{\alpha\beta}\right).  \label{02}
\end{equation}
% (02)
To obtain the electromagnetic $D$-momentum $P_\mu$, we must integrate
expression (\ref{02}) over a spacelike $D-1$-dimensional hypersurface.
However, straightforward integration is impossible because of divergences
pertaining to the singular behavior of $F_{\mu\nu}$. For instance, the $4D$
expression for $P_\mu$ diverges linearly because we have $%
F_{\mu\nu}\propto\rho^{-2}$ near the world line. We can eliminate this
divergence by renormalizing the mass of the charged particle \cite
{Teitelboim}.

For $4D$, the action for particles interacting with an electromagnetic field
is 
\begin{equation}
S_{{\rm p}}+S_{{\rm int}}=-\sum_{I=1}^K \int d\tau_I\,\Bigl[m^I_0\,\sqrt{{%
\dot z}_\alpha^I\,{\dot z}_I^\alpha}+ {\dot z}_\nu^I\,A^\nu(z_I)\Bigr],
\label{32}
\end{equation}
% (32)
where $m^I_0$ is the bare mass of the $I$th particle (all electric charges
of particles are set equal to unity) and $z^I_\alpha$ are the coordinates of
the $I$th particle. Expression (\ref{32}) is reparametrization invariant.
This is important for choosing the correct integration procedure for
singular $4D$ quantities. Therefore, it is natural to preserve this symmetry
in the general $D$-dimensional case.

We thus assume the $D$-dimensional action $S$ to be invariant w.\ r.\ t.\
the infinitesimal reparametrization transformations $\delta
\tau=\varepsilon,\,\delta z_\mu=v_\mu\varepsilon$. The second Noether
theorem then implies the identity 
\[
\frac{\delta S}{\delta z_\mu}\,v_\mu=0.
\]
Therefore, the Eulerian $\delta S/\delta z_\mu$, governing the dynamics of a
bare particle, contains the operator of projection on a $D-1$-dimensional
hyperplane with the normal vector $v^\mu$, 
\begin{equation}
\stackrel{\scriptstyle v}{\bot}\!(\,{\dot p}_{\hskip0.5mm 0}-f\,)=0,
\label{18}
\end{equation}
% (18)
where $p_{\hskip0.5mm 0}^{\hskip0.5mm\mu}$ is the $D$-momentum of the bare
particle and $f^\mu$ is the $D$-force that acts on this particle.

Equation (\ref{18}) is in fact the $D$-dimensional form of Newton's second
law. Indeed, the projection operator $\stackrel{\scriptstyle v}{\bot }$
embeds a $D-1$-dimensional Newton dynamics in spacelike hyperplanes, which
are ``common'' $D-1$-dimensional spaces in the instantly co-moving inertial
frame of reference. If the self-interaction is correctly taken into account,
the reparametrization invariance preserved, and the equation of motion of
the dressed particle has the same projection structure as the equation of
motion for the bare particle (\ref{18}). Nevertheless the dynamics of the
dressed and bare particles can differ because the dependences of the $D$%
-momenta $p^{\hskip0.5mm\mu }$ and $p_{\hskip0.5mm0}^{\hskip0.5mm\mu }$, on
kinematical variables can differ.

\subsubsection{The $2D$ self-interaction}

Substituting $F_{\mu\nu}$ (\ref{4a}) in the energy-momentum tensor (\ref{02}%
), we obtain 
\begin{equation}
\Theta_{\mu\nu}={\frac{1}{4}}\,(\,c_\mu v_\nu+v_\mu c_\nu-c_\mu c_\nu\,)= {%
\frac{1}{4}}\,(\,v_\mu v_\nu-u_\mu u_\nu\,)={\frac{1}{4}}\,\eta_{\mu\nu}.
\label{1-2}
\end{equation}
% (1-2)
We have used the completeness relation $\eta_{\mu\nu}= v_\mu v_\nu-u_\mu
u_\nu$ valid in the two-dimensional spacetime where $v^{\mu}$ and $u^{\mu}$
span a basis. It is clear from (\ref{1-2}) that $\partial_\mu
\Theta^{\mu\nu}=0$. Therefore, the integration hypersurface $\Sigma$ in the
expression for the two-momentum of the electromagnetic field $P_\mu$ is
arbitrary. In particular, we may take $\Sigma$ to be spacelike hyperplane (a
straight line) with the normal $v^{\mu}$. Then the integration is trivial
and results in the expression 
\begin{equation}
P_{\mu}=v_{\mu}\,{\frac{L}{2}}\,,\quad L\to\infty  \label{3-2}
\end{equation}
% (3-2)
which linearly diverges at large $L$.

If the action (\ref{32}) is used to describe the particle dynamics, this
divergence can be absorbed in the particle mass renormalization. Then the
two-momentum and the equation of motion of the dressed particle coincide
with the respective two-momentum and the equation of motion of the bare
particle with the observable mass $m$ in place of the bare mass $m_0$. The
only effect of the self-interaction in $2D$ is the mass renormalization. In
contrast to the case $4D$, where the mass renormalization eliminates an
ultraviolet divergence, the mass renormalization for $D=2$ eliminates the
infrared divergence.

It is interesting to compare an {\it effective} $2D$ case and a {\it genuine}
$2D$ case. Consider a particle that moves along a straight line in the $4D$
Minkowski spacetime, e.\ g., along the $x$ axis, and is subject to the
action of a force directed along the same line. If the projection structure
of Eq.\ (\ref{18}) is taken into account by the relations 
\begin{equation}
v^\mu=(\cosh\alpha,\,\sinh\alpha,\,0,\,0),\quad
f^\mu=f\,(\sinh\alpha,\,\cosh\alpha,\,0,\,0),  \label{1d}
\end{equation}
% (1d)
then the equation of motion for the dressed particle (the Lorentz--Dirac
equation) becomes 
\[
m{\dot\alpha}-\frac23\,{\ddot\alpha}=f.
\]
On the other hand, the equation of motion for the dressed particle in the
genuine $2D$ world does not contain the term of the radiation damping, 
\[
m{\dot v}=f.
\]
The reason for this difference is that in the $3D$ space, only the
kinematics and interparticle interactions can be effectively
one-dimensional; the self-interaction remains three-dimen\-sional because
its form is fixed by the pole singularity of the three-dimensional Coulomb
potential. Does it means that electromagnetic energy is dot radiated in the $%
2D$ world? The general definition \cite{Teitelboim,kosy} states that
radiation is the part of the energy-momentum density that

(i) leaves the source at the speed of light,

(ii) does not dynamically depend on the remaining part of the
energy-momentum field tensor outside the world line, and

(iii) produces a constant flux when integrated over any closed surface
enclosing the source.

This definition can be formulated more exactly if we split the
energy-momentum tensor of electromagnetic field in two parts, $\Theta _{\mu
\nu }=\Theta _{\mu \nu }^{{\rm I}}+\Theta _{\mu \nu }^{{\rm II}}$, and
interpret the term $\Theta _{\mu \nu }^{{\rm II}}$ as the radiation term.

Because a part of the energy-momentum tensor propagates at the speed of
light, this part runs along rays of the forward light cone. In other words,
condition (i) implies that the term $\Theta_{\mu\nu}^{{\rm II}}$ contributes
zero to the flux through the forward light cone. Because the normal vector
to the forward light cone with the vertex at a point on the world line is
the vector $c_\mu$ drawn from this point, the flux of the quantity $%
\Theta_{\mu\nu}^{{\rm II}}$ through such a hypersurface vanishes if 
\begin{equation}
\Theta_{\mu\nu}^{{\rm II}}\propto c_\mu c_\nu.  \label{i}
\end{equation}
% (i)
Thus, relation (\ref{i}) is equivalent to condition (i).

The local energy-momentum conservation law $\partial^\mu \Theta_{\mu\nu}=0$
is satisfied outside world lines. This law can be interpreted as the mutual
dynamic independence of electromagnetic and mechanical energy-momentum parts
in the domain under consideration. If, moreover, the equality 
\begin{equation}
\partial^\mu \Theta^{{\rm II}}_{\mu\nu}=0  \label{ii}
\end{equation}
% (ii)
holds outside world lines, then the term $\Theta_{\mu\nu}^{{\rm II}}$ may be
taken to be dynamically independent of $\Theta_{\mu\nu}^{{\rm I}}$. Relation
(\ref{ii}) expresses condition (ii).

A spatial decrease of $\Theta_{\mu\nu}^{{\rm II}}$ must be related to an
increase in the area of the sphere enclosing the source; the relation is
such that the energy flux through the sphere remains constant for any
radius, 
\begin{equation}
\Theta_{\mu\nu}^{{\rm II}}\propto \rho^{-(D-2)}.  \label{iii}
\end{equation}
% (iii)
Relation (\ref{iii}) expresses condition (iii). It is commonly assumed that $%
\Theta_{\mu\nu}^{{\rm I}}$ decreases faster than $\Theta_{\mu\nu}^{{\rm II}}$%
, so that radiation becomes separated from the rest of the rest of the
electromagnetic energy-momentum tensor when the source is asymptotically far
away. Nevertheless, we explicitly impose the (additional) condition that the
asymptotic behavior of $\Theta_{\mu\nu}^{{\rm I}}$ and $\Theta_{\mu\nu}^{%
{\rm II}}$ are different.

In expression (\ref{1-2}) we segregate the term $-c_\mu c_\nu/4$, which
satisfy conditions (\ref{i})--(\ref{iii}). However, its spatial behavior is
similar to that of other terms from expression (\ref{1-2}), i.\ e., it does
not depend on $\rho$. Each such term generates an energy flux through the
zero-dimensional sphere (two points on the spacelike straight line that lie
in opposite directions at equal distances from the source) and this flux is
not altered with distance from the source. In the case $D=2$, it is
impossible to segregate the term $\Theta_{\mu\nu}^{{\rm II}}$ satisfying all
conditions (\ref{i})--(\ref{iii}) whereas the term $\Theta_{\mu\nu}^{{\rm I}}
$ decreases with distance from the source. In other words, there is no
radiation for $D=2$. Note that in the $2D$ case, the electric field $F_{01}$
is present but the magnetic field is absent; the Pointing vector is
therefore identically zero. The energy dissipation is therefore absent in
the $2D$ picture, and all particle motions are reversible.

\subsubsection{The $6D$ self-interaction}

The field $F_{\mu\nu}$ generated by a single particle is given by expression
(\ref{4}). Substituting (\ref{4}) in (\ref{02}), we obtain the first term of 
$\Theta_{\mu\nu}$, 
\[
F_{\mu\alpha}F^\alpha_{\hskip1.5mm\nu}=\frac{1}{9}\,\Biggl\{
-\frac{1}{\rho^6}\,\left[(a_\mu a_\nu+a^2\,v_\mu v_\nu)+(c\cdot V)\, (c_\mu
V_\nu+c_\nu V_\mu)-c_\mu c_\nu\,V^2\right]+ 
\]
\begin{equation}
+\frac{1}{\rho^3}\,\biggl[\,a_\mu V_\nu+a_\nu V_\mu +(a\cdot V)\,(v_\mu
c_\nu+v_\nu c_\mu) -(v\cdot V)\,(a_\mu c_\nu+a_\nu c_\mu) -\frac{\lambda+1}{%
\rho}\,(v_\nu V_\mu+v_\mu V_\nu)\biggr]\Biggr\}.  \label{108}
\end{equation}
%(108)
It follows 
\begin{equation}
F_{\alpha\beta}F^{\alpha\beta}=\frac{2}{9}\,\left[\, \frac{a^2}{\rho^6}%
-(c\cdot V)^2-\frac{2}{\rho^3}\,(a\cdot V)+ \frac{2}{\rho^3}\,\frac{\lambda+1%
}{\rho}\,(v\cdot V)\right].  \label{109}
\end{equation}
%(109)
Integration of (\ref{108}) and (\ref{109}) over $\rho^4 d\rho$ results in
cubic and linear divergences. The former can be absorbed in the
renormalization of the particle mass, and for the latter to be absorbed the
Lagrangian should contain physical parameters of the appropriate dimension.

We add the reparametrization invariant term 
\begin{equation}
\sum_{I=1}^K\kappa_0^I \int d\tau_I\, \gamma_I\left(\frac{d}{d\tau_I}\,%
\biggl({\gamma_I}\, \frac{dx^\mu_I}{d\tau_I}\biggr)\right)^2, \quad
\gamma_I\equiv\biggl({\frac{dx_\nu^I}{d\tau_I}\,\frac{dx^\nu_I}{d\tau_I}} %
\biggr)^{-1/2}  \label{222}
\end{equation}
% (222)
to the action (\ref{32}). The six-momentum of the bare particle then becomes 
\begin{equation}
p_{\hskip0.5mm 0}^{\hskip0.5mm\mu}=m_0\,v^\mu+ \kappa_0\,(2\,{\dot a}%
^\mu+3\,a^2\,v^\mu),  \label{322}
\end{equation}
% (322)
whence the dimensionality considerations show that the cubic and linear
divergences are absorbed by the respective renormalization of $m_0$ and $%
\kappa_0$. Thus, passing from $4D$ to higher dimensions (with the
understanding that all ultraviolet divergences would be absorbed in
renormalization of physical quantities) implies the appearance of higher
derivatives in the particle Lagrangian. Hence, a self-consistent field
theory in a spacetime of dimension $D>4$ is possible only if the particle
behavior is governed by a rigid dynamics.

A customary procedure for finding the equation of motion for a dressed
particle is to find a regularization that preserves the reparametrization
invariance of the action with a subsequent calculation of all divergent
integrals and a segregating from them of the parts that are regular when
eliminating the regularization. This is a rather laborious problem, which
can be solved, however, in a simpler way. In this approach, it is enough to
find the radiation rate (which is expressed by a convergent integral, i.\
e., is independent of the regularization scheme), and impose the condition
that the equation of motion for a dressed particle be endowed with the
projection structure.

We now find the expression for the radiation rate. A $4D$ closed surface
enclosing the source plays the same role in the $6D$ case as a $2D$ surface
in the $4D$ case. From (\ref{108}) and (\ref{109}) follows that the only
summand containing the terms that generate the energy-momentum flux, which
is constant at any distance from the source, is $-V^2\,c_\mu\,c_\nu$. In
(\ref{5}), we segregate the term proportional to $\rho^{-2}$: 
\begin{equation}
b^\mu =\frac{1}{3}\left(\frac{{\dot a}^\mu}{\rho^2}-3\,\frac{\lambda+1} {%
\rho^3}\,a^\mu+3\biggl(\frac{\lambda+1}{\rho^2}\biggr)^2\,v^\mu- \frac{({%
\dot a}\cdot c)\,v^\mu}{\rho^2}\right).  
\label
{b-mu-df}
\end{equation}
We can easily verify that the tensor $b^2 c_\mu c_\nu$ meets all the
conditions (\ref{i})--(\ref{iii}) and is therefore the radiation term: $%
\Theta_{\mu\nu}^{{\bf II}}=b^2 c_\mu c_\nu$.

In view of (\ref{ii}), the integration hypersurface in the definition of the
radiated six-momentum $P^{\mu}_{{\rm rad}}$ may be chosen arbitrarily. It is
convenient to integrate over a timelike five-dimensional tube ${\cal T}$ of
a small radius $\rho=\epsilon$ enveloping the source world line. The area
measure element of this tube is 
\begin{equation}
d\sigma^\mu=\partial^\mu\!\rho\,\rho^4\,d\Omega_4\,d\tau=(v^\mu+ \lambda
c^\mu)\,\epsilon^4\,d\Omega_4\,d\tau,  \label{tube-measure}
\end{equation}
and the problem is reduced to calculating the integral 
\[
-\frac{1}{9}\int d\tau\int d\Omega_4\,\Biggl\{\biggl[(\stackrel{\scriptstyle %
v}{\bot}{\dot a})^2+9(a\cdot u)^2a^2+9(a\cdot u)^4+({\dot a}\cdot u)^2%
\biggr]
v^\mu-\biggl[3(a\cdot u) (a^2)^{.}+6(a\cdot u)^2({\dot a}\cdot u)\biggr]
u^\mu\Biggr\}
\]
where $(a^2)^{.}\equiv da^2/d\tau$. Using the relations 
\[
\int d\Omega_{4}\, u_\mu u_\nu=-\frac{\Omega_{4}}{5}\, 
\stackrel{\scriptstyle v}{\bot}_{\hskip0.5mm\mu\nu},
\]
\[
\int d\Omega_{4}\,u_\alpha u_\beta u_\mu u_\nu=\frac{\Omega_{4}}{5\cdot 7}%
\,\left( \stackrel{\scriptstyle v}{\bot}_{\hskip0.5mm\mu\nu}\, \stackrel{%
\scriptstyle v}{\bot}_{\hskip0.5mm\alpha\beta}+ \stackrel{\scriptstyle v}{%
\bot}_{\hskip0.5mm\alpha\mu}\, \stackrel{\scriptstyle v}{\bot}_{\hskip%
0.5mm\beta\nu}+ \stackrel{\scriptstyle v}{\bot}_{\hskip0.5mm\alpha\nu}\, 
\stackrel{\scriptstyle v}{\bot}_{\hskip0.5mm\beta\mu}\right) 
\]
we find 
\begin{equation}
{P}^{\mu}_{{\rm rad}}= \int_{{\cal T}} d\sigma_\nu \Theta^{\mu\nu}_{{\bf II}%
}= \frac{1}{9} \int_{-\infty}^\tau ds\,\Biggl[-\frac{4}{5}\biggl({\dot a}^2-%
\frac{16}{7}\, (a^2)^2 \biggr)v^\mu-\frac{3}{7}\,(a^2)^{.}\,a^\mu +\frac{6}{%
5\cdot 7}\,a^2 (\stackrel{\scriptstyle v}{\bot}{\dot a})^\mu\Biggr].
\label{06}
\end{equation}

Integration of the remainder of the electromagnetic energy-momentum tensor
results in the bound six-momentum containing both divergent and finite
terms. The finite term $p_{\hskip0.5mm{\rm fin}}^{\mu}$ has the same
dimension as the radiated six-momentum ${P}^{\mu}_{\hskip0.3mm{\rm rad}}$,
and hence the same dependence on kinematic variables: 
\[
p_{\hskip0.5mm{\rm fin}}^{\mu}=c_1\,{\ddot a}^{\mu}+
c_2\,a^2a^\mu+c_3\,(a^2)^{.}\,v^\mu
\]
where $c_1$, $c_2$ and $c_3$ are numerical coefficients. Since we require
the equation of motion for a dressed particle be endowed with the projection
structure, the sum of projections of the two quantities ${\dot p}_{\hskip%
0.5mm{\rm fin}}^{\mu}$ and ${\dot P}^{\mu}_{{\rm rad}}$ (which have the same
dimension) on the direction $v^\mu$ must be zero \footnote{%
We here use the reparametrization invariance requirement embodied in the
projection structure of the renormalized equation of motion.}. Hence, using
the identities 
\[
(a\cdot v)=0, \quad (\dot{a}\cdot v)=-a^2, \quad (\ddot{a}\cdot v)=-\frac{3}{%
2}\,(a^2)^{.}, \quad (\stackrel{\ldots}{a}\cdot\,v)=-2\,(a^2)^{.\,.}+\dot{a}%
^2,
\]
we find 
\begin{equation}
p_{\hskip0.5mm{\rm fin}}^{\mu}=\frac{4}{45}\,\biggl({\ddot a}^{\mu}+ \frac{16%
}{7}\,a^2\,a^\mu+2(a^2)^{.}\,v^\mu \biggr).  \label{05}
\end{equation}
% (05)

To find the divergent term, the integration is most conveniently performed
over the surface of the forward light cone. The kinematic structure of the
divergent term is identical to that of  the bare particle six-momentum (\ref
{322}). Adding them, we have a finite quantity 
\begin{equation}
{\bar p}^{\mu}=mv^\mu+\kappa\,(2{\dot a}^\mu+3a^2v^\mu),  \label{bar-p}
\end{equation}
% (bar-p)
where $m$ and $\kappa$ are the renormalized parameters corresponding to the
respective bare parameters $m_0$ and $\kappa_0$ in (\ref{322}). Adding ${%
\bar p}^{\mu}$ and $p_{\hskip0.5mm{\rm fin}}^{\mu}$, we obtain the total
bound six-momentum 
\begin{equation}
{\pi}^{\mu}=mv^\mu+\kappa\,(2{\dot a}^\mu+ 3a^2\,v^\mu)+\frac{4}{45}\,\biggl(%
{\ddot a}^{\mu}+\frac{16}{7}\, a^2a^\mu+2(a^2)^{.}\,v^\mu\biggr).  \label{pi}
\end{equation}

Integration of the tensor $\Theta _{\hskip0.3mm{\rm mix}}^{\mu \nu }$ formed
from the mixed contribution of an external field $F_{\hskip0.3mm{\rm ex}%
}^{\mu \nu }$ and the retarded field of the given particle $F^{\mu \nu }$
results in a finite six-momentum $P_{\hskip0.3mmmix}^{\hskip0.3mm\mu }$. It
can be performed over a tube ${\cal T}$ of an infinitesimal radius
enveloping the source world line with the use of the measure $d\sigma ^{\mu }
$ of the form (\ref{tube-measure}). As (\ref{4}) and (\ref{5}) suggest, only
the most singular terms of $F^{\mu \nu }$ behaving as $\rho ^{-4}$
contribute to the integral: 
\begin{equation}
P_{\hskip0.3mm{\rm mix}}^{\mu }=\int_{{\cal T}}d\sigma _{\nu }\Theta _{{\rm %
mix}}^{\mu \nu }=-\int_{-\infty }^{\tau }ds\,F_{\hskip0.3mm{\rm ex}}^{\mu
\nu }\,v_{\nu }.  \label{P-ex-6D}
\end{equation}
The quantity ${\dot{P}}_{\hskip0.3mm{\rm mix}}^{\mu }$ is therefore the
negation of the external Lorentz six-force $f^{\mu }=F_{\hskip0.3mm{\rm ex}%
}^{\mu \nu }\,v_{\nu }$.

We now write the balance of six-momenta for the system as a whole 
\begin{equation}
\dot{\pi}^{\mu}+{\dot P}^{\mu}_{\hskip0.3mm{\rm rad}}+ {\dot P}^{\mu}_{\hskip%
0.3mm{\rm mix}}=0.  \label{balance-p-6D}
\end{equation}
% (04)
Substituting (\ref{pi}), (\ref{06}), and (\ref{P-ex-6D}) in this equation,
we obtain the equation of motion for the dressed particle 
\begin{equation}
\stackrel{\scriptstyle v}{\bot}\!(\,{\dot p}-f)=0  \label{eq-motion-dress-6d}
\end{equation}
% (n)
where the dressed particle six-momentum $p^\mu$ is defined as 
\begin{equation}
p^\mu=mv^\mu+\kappa\,(2{\dot a}^\mu+3a^2 v^\mu)+ \frac{1}{9}\,\biggl(\frac{4%
}{5}\,\ddot{a}^\mu+2a^2a^\mu+ (a^2)^{.}\,v^\mu\biggr).  
\label
{p-mu-6D}
\end{equation}

In conclusion, we discuss the problem of the Zitterbewegung of a charged
particle. A free bare particle with the six-momentum (\ref{322}) may not
only execute a uniform rectilinear Galilean motion but also tremble. To see
this, we consider the rectilinear motion \footnote{%
More general Zitterbewegungs of a free rigid particle were considered in 
\cite{kn}}. The projection structure of the equation of motion is then taken
into account using the relations (\ref{1d}), and we get 
\[
m_{0}\,{\dot{\alpha}}+\kappa _{0}\,(2\stackrel{\ldots }{\alpha }-{\dot{\alpha%
}}^{3})=0.
\]
If $m_{0}$ and $\kappa _{0}$ have the same sign, then the solution
oscillates in the domain of small accelerations ($|{\dot{\alpha}}|<\sqrt{%
m_{0}/\kappa _{0}}$).

Electromagnetic self-interaction essentially affects the free propagation
regime because the equation of motion is modified by higher derivative
terms. Are there Zitterbewegungs that are not accompanied by radiation? Such
a non-radiation regime would in principle be possible if the expression for
the radiation rate would contain only derivatives of powers that are higher
than the highest power of derivatives entering the equation of motion for
the bare particle. However, expression (\ref{06}) shows that the $6D$
radiation rate depends on $a_{\mu }$ and $\dot{a}_{\mu }$, and hence each
Zitterbewegung, being an accelerated motion, is accompanied by radiation.
This results in damping the Zitterbewegung and obtaining the asymptotic
regime of Galilean uniform rectilinear motion.

\section{The Yang--Mills--Wong theory}

Consider a system of $K$ classical pointlike particles interacting with the
gauge SU$(N)$ Yang--Mills field. Particles, which are enumerated by the
index $I$, $I=1,\ldots ,K$, have color charges $Q_{I}^{a}$ belonging to
the conjugate representation of the group SU$(N)$; the color index $a$
ranges over integers from $1$ to $N^{2}-1$. 
The Yang--Mills--Wong $D$-dimensional action is \footnote{The four-dimensional 
Yang--Mills--Wong theory is discussed, e. g., in \cite{bbs, k}.} 
\begin{equation}
S=S_{{\rm p}}-\sum_{I=1}^{K}\,\int d\tau _{I}\,{\rm tr}\,\lambda_{I}^{-1}
Z_{I}{\dot{\lambda}}_{I}-\int d^{D}x\,{\rm tr}\left(j_\mu\,A^\mu+
\frac{1}{4\,\Omega _{D-2}}\,F_{\mu\nu}\,F^{\mu\nu}\right) ,
\label{act}
\end{equation}
where $S_{{\rm p}}$ depends on kinematic variables in the same way as the
corresponding action term of the $D$-dimensional electrodynamics, 
$\lambda_{I}=\lambda _{I}(\tau _{I})$ are time-dependent elements of the 
algebra su$(N)$, and $Z_{I}$ are combinations of the su$(N)$-algebra 
generators $T_{a}$, $Z_{I}=e_{I}^{a}T_{a}$ with constant $e_{I}^{a}$. 
These combinations determine the color charges of particles, 
$Q_{I}=\lambda _{I}Z_{I}\lambda_{I}^{-1}$ \cite{bbs}. 
The color current of particles is 
\[
j_{\mu }(x)=\sum_{I=1}^{K}\int\!d\tau_{I}\,Q_{I}(\tau_{I})\,v_{\mu}^{I}
(\tau_{I})\,\delta^{D}\Bigl(x-z_{I}(\tau _{I})\Bigr) ,
\]
where $Q_{I}=Q_{I}^{a}\,T_{a}$. The field strength is standard, 
\[
F_{\mu \nu }=\partial _{\mu }A_{\nu }-\partial _{\nu }A_{\mu }-ig\,\lbrack
A_{\mu },\,A_{\nu }\rbrack ,
\]
where $g$ is the coupling constant. Electrodynamics is a particular case
with the Abelian gauge group U(1).

Action (\ref{act}) produces the Yang--Mills equations 
\begin{equation}
\Box A_\mu-\partial_\mu \partial_\nu A^\nu -ig\,\Bigl(\partial_\nu[%
A^\nu,A_\mu]+[A^\nu,\partial_\mu A_\nu-\partial_\nu A_\mu]\Bigr) +
g^2\bigl[A_\nu,[A^\nu,A_\mu] \bigr]=\Omega_{D-2}\, j_\mu  
\label{ym}
\end{equation}
and the Wong equations 
\begin{equation}
\dot Q_I=-ig\,[Q_I,\,v^I_\mu\,A^\mu (z_I)],  
\label{w}
\end{equation}
which describe the evolution of color charges $Q_I$.

We must verify that the theory with the action (\ref{act}) is consistent in
the sense that the dimensional parameters entering $S_{{\rm p}}$ suffice for
absorbing all ultraviolet divergences.

Exact retarded solutions of $4D$ equations (\ref{ym}) and (\ref{w}) were
considered in detail in \cite{k}. They were obtained using the Ansatz 
\[
A_\mu(x)=\sum^K_{I=1}\,\sum^{N^2 -1}_{a=1} T_a\,(v_\mu^I\,
f^{aI}+R_\mu^I\,h^{aI}), 
\]
where $f^{aI}(x)$ and $h^{aI}(x)$ are the desired functions. This Ansatz is
a non-Abelian generalization of expression (\ref{30}). 
The transition to $6D$ and $2D$ can be performed similar to the above case
where we obtained expressions (\ref{3}) and (\ref{3a}) as extensions of the 
Ansatz (\ref{30}) in the electrodynamics.

However, if we are interested in the singular behavior of the Yang--Mills
field at small distances from the source, we do not need to solve 
Eqs.\ (\ref{ym}) and (\ref{w}), it suffices to use the Gauss law. 
Indeed, when $\rho\to 0$, the area of the surface enclosing the particle 
decreases as $\rho^{D-2}$, and the volume as $\rho^{D-1}$; therefore, the 
color charge of the field,
which is proportional to the volume, asymptotically decreases, and the total
color charge tends to the color charge of the pointlike particle, which is
expressed as the flux of the Li{\'e}nard--Wiechert field strength 
$F_{\mu\nu}$ through the given surface 
\footnote{The Gauss law for $D=4$ was discussed in detail in \cite{kosy}.}. The
pointlike particle charge $Q_I$ therefore makes the leading contribution to
the field strength at short distances from this particle; moreover, 
\begin{equation}
F_{\mu\nu}\sim \frac{Q_I}{\rho^{D-2}},\quad \rho\to 0.  \label{g}
\end{equation}
The leading Yang--Mills field singularity is the same as in the case of
electromagnetic field; hence, the action term ${S}_{{\rm p}}$ is the same in
both these theories.

Equation (\ref{g}) implies that the vector potential $A_\mu$ is proportional
to $Q_I$ and tends asymptotically to $\rho^{3-D}$ if $D>4$. Substituting
such an $A_\mu$ in the right-hand side of Eq.\ (\ref{w}) yields zero; hence $%
\dot Q_I=0$. Therefore, the singular behavior of $A_\mu$ is severely
restricted in the presence of $\delta$-function-like colored objects, which
in turn makes the color charge evolution of such objects trivial \footnote{%
The presence of a $\delta$-function-like distribution of color charge
forbids an explicit interwining of color and spacetime indices of $A_\mu$,
which occurs in monopole or instantonic solutions.}.

If $h^{aI}=0$, then the solution becomes Abelian and Coulomb-like in the
total spacetime. 
For $D=4$, non-Abelian solutions is known to exist \cite{k}. 
However, for $D=2$ where $f^{aI}=0$, non-Abelian solutions are absent. 
The question arises whether non-Abelian solutions exist for any dimension 
$D>4$?

We now show that $6D$ field configurations generated by a colored particle
in the case of the SU(2) gauge group can only be Abelian. We combine the
Pauli matrices $\sigma_{\pm}=\sigma_{1}\pm\sigma_{2}$ and $\sigma_{3}$, and 
the vector potential becomes 
$A=\alpha_{3}\sigma_{3}+\alpha_{+}\sigma_{+}+\alpha_{-}\sigma_{-}$. 
Hence, the terms of the second and third
orders in fields in Eq.\ (\ref{ym}) have the respective symbolic forms 
\[
\partial \,\lbrack A,A\rbrack =\partial \,(\alpha _{+}\alpha _{-})\,\sigma
_{3}+\partial \,(\alpha _{+}\alpha _{3})\,\sigma _{+}+\partial \,(\alpha
_{-}\alpha _{3})\,\sigma _{-},
\]
\[
\lbrack A,\lbrack A,A\rbrack \rbrack =(\alpha _{-}\alpha _{+}\alpha
_{3})\,\sigma _{3}+(\alpha _{+}\alpha _{3}\alpha _{3})\,\sigma _{+}+(\alpha
_{-}\alpha _{3}\alpha _{3})\,\sigma _{-}.
\]
Let the color charge of a particle be proportional to $\sigma _{3}$. By
virtue of the Gauss law at small distances, the quantity $\alpha _{3}$ is
proportional to $\rho ^{-3}$ as $\rho \rightarrow 0$, and its second
derivative behaves as $\rho ^{-5}$. Supposing that $\alpha _{+}$ and $\alpha
_{-}$ are nonzero, we conclude that the terms of the second and third orders
in Eq.\ (\ref{ym}) must have the same asymptotic behavior. Then, comparing
the coefficients of $\sigma _{3}$ in the terms of second and third orders,
we conclude that the quantity $\alpha _{-}\alpha _{+}$ must have the
respective asymptotic behaviors $\rho ^{-4}$ and $\rho ^{-2}$. This means
that at least one of the quantities $\alpha _{-}$ or $\alpha _{+}$ must be
zero. Setting $\alpha_{-}=0$ and taking into account that differentiating 
$\alpha_{+}$ raises its singularity power by one, we find that the
coefficient of $\sigma _{+}$ in terms of the first, second, and third orders
in Eq.\ (\ref{ym}), must have the respective singularity powers 2,4, and 6.
Therefore, $\alpha _{-}=\alpha _{+}=0$, and the vector potential $A_{\mu }$
must be Abelian.

The general reason why a pointlike colored particle may generate a
non-Abelian field only in the four-dimensional spacetime is as follows. The
Gauss law in $D=4$ dictates that a vector potential must have only a simple
pole, $A_{\mu }\propto \rho ^{-1}$. Then, both differentiation and
multiplication by $A_{\mu }$ raise the singularity power by one. Therefore,
acting with the derivative $\partial _{\mu }$ or the covariant derivative $%
\partial _{\mu }-igA_{\mu }$ on singular terms of the vector potential $%
A_{\mu }$, we obtain the same asymptotic behavior. Because of this,
non-Abelian solutions are admitted.

Therefore, the set of admitted field configurations of the Yang--Mills--Wong
theory in the spacetime of any even dimension $D=2n$ except $D=4$ is
equivalent to the corresponding field  configuration set of the 
$D$-dimensional electrodynamics. 
\vskip5mm \noindent {\large {\bf Acknowledgments}} \vskip2mm \noindent 
The work was supported in part by the
International Science and Technology Center under Project {\#} 208.
\vskip5mm \noindent {\large {\bf Note added}} \vskip2mm \noindent 
Compared to the 1999 paper, the normalization of $6D$ vector potential
in (\ref{36}) is corrected.
(The inaccuracy in the normalization was noticed in \cite{kls}.)
This results in corrections of overall coefficients in expressions
(\ref{4}), (\ref{108}), (\ref{109}), (\ref{b-mu-df}), (\ref{06}),  
and overall coefficients of last terms in 
(\ref{pi}) and  (\ref{p-mu-6D}).

\end{document}